\documentstyle[twoside,fleqn,espcrc2]{article}
\def\epsfpreprint{Y}   

\if \epsfpreprint Y \input epsf \fi
\def\half{{1 \over2}} 
\def\sla{\raise.15ex\hbox{$/$}\kern-.57em}

\newcommand{\AmS}{{\protect\the\textfont2
  A\kern-.1667em\lower.5ex\hbox{M}\kern-.125emS}}

\def\captionA{
The nonperturbatively determined $\beta$--function for the Yukawa coupling
at $N_c ,N_w \rightarrow\infty$ with
${{N_c}\over{N_w}}={{3}\over {2}}$.
At tree level the top quark mass is given by
$m_{t }=(4\pi^2  \alpha /3)^{\half} ~ 246~[GeV]:~\beta (\alpha ) = {{d\alpha}
\over {dt}}$
where $t$ is the logarithm of the energy scale.
}

\title{A better large $N$ expansion for chiral Yukawa models.\thanks
{This research was supported in part by the DOE under grant
\# DE-FG05-90ER40559.}}
\author{George Bathas\address{Department of Physics and Astronomy\\
Rutgers University, Piscataway, NJ 08855, U.S.A.}
and Herbert Neuberger
$^{\rm a}$ \thanks{Speaker at the Conference.
}}

\begin{document}

\begin{abstract}
We consider the most general renormalizable chiral Yukawa model with
$SU(3)_{\rm color}$ replaced by $SU(N_c)$, $SU(2)_{\rm L}$ replaced
by $SU(N_w )$ and $U(1)_{Y}$ replaced by
$U(1)^{N_w -1}$ in the limit
$N_c \rightarrow\infty$, $N_w \rightarrow\infty$
with the ratio $\rho=\sqrt{{N_w}\over{N_c}} \ne 0,\infty$ held fixed.
Since for $N_w \ge 3$ only one renormalizable Yukawa coupling per
family exists and there is no mixing between families
the limit is appropriate for
the description of the effects of a heavy top quark when
all the other fermions are taken to be massless.  The large $N=\sqrt{N_{c}
N_{w}}$ expansion
is expected to be no worse quantitatively in this model that in the purely
scalar case and the $N=\infty$ limit is soluble even when the model is
regularized non--perturbatively. A rough
estimate of the triviality bound on the Yukawa coupling is equivalent to $m_t
\le 1~TeV$.

\end{abstract}

\maketitle
\addtocounter{footnote}{-2}

 We wish to study non--perturbatively a model that is reasonably close to what
the minimal
standard model would look like if the top quark Yukawa coupling and the Higgs
selfcoupling
were both large while all other couplings were kept at their known small values
and could
therefore be neglected in a first approximation.

We propose to do this using an ${1\over N}$ expansion on a model defined
non--perturbatively with
a Pauli Villars type of regularization\cite{paper}. The action is:

\begin{eqnarray}
-S=\int_x  \bar \psi_{-}^{ia}\sla\partial_{+}\psi_{+}^{ia}+
\bar\chi_{+}^{a}h_{\chi}(-\partial^2 )\sla\partial_{-}\chi_{-}^{a}
\cr+
\phi^{i*}h_{\phi}(-\partial^2 )\partial^2 \phi^i -m_0^2 \phi^{i*}\phi^i -
{{\lambda_0}\over{4N_w}} (\phi^{i*}\phi^{i})^2
\cr +{g_0 \over{\sqrt{N}}}
[
\bar\psi_{-}^{ia}\chi_{-}^{a}\phi^{i}+
\bar\chi_{+}^{a}\psi_{+}^{ia}\phi^{i}];~~~g_0 > 0
\label{eq:act}
\end{eqnarray}
where
\begin{eqnarray}
N=\sqrt{N_w N_c }, ~~\sla\partial_{+}=\sigma_{\mu}\partial_{\mu},~~
\sla\partial_{-}=\bar\sigma_{\mu}\partial_{\mu}\cr
\bar\sigma_{1,2,3} =\sigma_{1,2,3} ,~~
\bar\sigma_4 =-\sigma_4 =-i\cr
i=1,\cdots ,N_w~~a=1,\cdots,N_c
\label{eq:defs}
\end{eqnarray}
The $\sigma_{1,2,3}$ are Pauli matrices.
Ultraviolet divergences can be eliminated by just regularizing the
$\chi$ and $\phi$ propagators because there are no loops consisting of
only $\psi$ propagators. This is achieved with
$h_{\chi}(p^2 )=h_{\phi}(p^2 )=1+({p^2}/{\Lambda^2})^n$
with $n \ge 3$.

$\psi$ represents the top--bottom doublet, $\chi$ the right handed top and
$\phi$ the Higgs doublet.
The index $i$ generalizes the weak isospin index and the index $a$ labels
colors. $\psi_{+}$ is
a Euclidean left handed Weyl spinor, $\chi_{-}$ a right handed one and $\phi$ a
complex scalar.
$\bar  \psi_{-}$ and $\bar \chi_{+}$ are independent fields of handedness
opposite to their unbarred
counterparts as usual in the Euclidean functional integral formulation.

The large $N$ limit is taken with $\bar\rho\equiv\sqrt{{N_w}\over{2N_c}}$
held fixed and is dominated by planar diagrams. Renormalizability
allows us to get away with not having to introduce four Fermion
interactions and with them an insoluble planar diagram problem,
at no cost to generality.\footnote{Except for possible
cutoff effects.} Thus the model is manageable being similar
diagrammatically
to two dimensional QCD with matter at large $N_c$. The large $N$ limit
can be found by looking at the structure of
planar Feynman graphs: essentially one sums over all
``cactuses of bubbles'' and all ``rainbows''. For future work it
is somewhat preferable to use functional integral manipulations to
the same end.

As an example of what can be done in this scheme we present a very simple
calculation: We compute a $\beta$--function associated
with the Yukawa coupling constant defined in some simple manner and ignoring
cutoff effects. This can be done with relative ease because even
at infinite order in the couplings the flow of this
Yukawa coupling is unaffected
by $\lambda_0$, similarly to the finite $N$ form at one loop order. From
this calculation a rough estimate for the ``triviality'' bound on the
top mass can be obtained.\footnote{In practice this bound might be
somewhat uninteresting because it is likely to be higher than the
vacuum stability bound for reasonable Higgs masses.}

To find $\beta$ functions it suffices to solve the model at criticality
when all masses vanish. We start in the symmetric phase and pick a mass
independent renormalization scheme to make the zero mass limit smooth.

We introduce bilocal fields,
\begin{eqnarray}
S_{\alpha ,\beta}(x,y)
={1\over{N_c}} \bar\chi_{+\alpha}^a (x)\chi_{-\beta}^a (y)\cr
K(x,y)={1\over{N_w}} \phi^{i*}(x)\phi^{i}(y)
\label{eq:bilocals}
\end{eqnarray}
where $\alpha ,\beta$ are space--time spinor indices, into the functional
integral by writing representations for the appropriate $\delta$--functions
with the help of two auxiliary bilocal fields
$\lambda_{\alpha ,\beta} (x,y)$ and $\mu (x,y)$.
We can integrate out now all the original
fields, including the zero mode of the scalar field because we assumed that
we are in the symmetric phase. This makes the $N$ and $\bar\rho$ dependence
explicit and, when $N\rightarrow\infty$, the integral over
$S,\lambda ,K$ and $\mu$
is dominated by a saddle point which has to satisfy the following equations:
\begin{eqnarray}
K(x,y)=\{ [-h_{\phi}(-\partial^2 )\partial^2 + m_0^2 +\mu ]^{-1}\}_{y,x}\cr
S_{\alpha ,\beta}(x,y)=\{ [h_{\chi}(-\partial^2 )\sla\partial_{-} +
\lambda ]^{-1}\}_{y\beta ,x\alpha}\cr
\lambda_{\alpha ,\beta}(x,y)=-g^2 \bar\rho \sqrt{2}
\{\sla\partial_{+}^{-1}\}_{x\alpha ,y\beta } K(x,y)\cr
\mu (x,y) ={{g^2}\over{\bar\rho\sqrt{2}}}\sum_{\alpha ,\beta}
\{\sla\partial_{+}^{-1}\}_{x\alpha ,y\beta } S_{\alpha ,\beta} (x,y)
\cr+{{\lambda_0}
\over {2}} K(x,x)\delta_{x,y}
\label{eq:sp}
\end{eqnarray}
We go to Fourier space and use translational and rotational invariance to write
\begin{equation}
\tilde\lambda_{\alpha ,\beta} (p) =ip_{\mu}\bar\sigma_{\mu} \hat\lambda (p^2)~
{}~~~~~~~~\tilde\mu (p) =\hat\mu ( p^2 )
\label{eq:rotinv}
\end{equation}
Let us introduce $Z_{\chi}$ and $Z_{\phi}$, wave function
renormalization constants, and new functions $f(p^2 )=Z_{\chi} [\hat\lambda
(p^2 )
+h_{\chi} (p^2 )]$ and $g(p^2 )=Z_{\phi } [p^2 h_{\phi} (p^2 ) + m_0^2 +
\hat\mu (p^2 )]$. Since at the planar level there are no vertex corrections
to the Yukawa coupling\footnote{This is also true at
one loop order at finite $N$.} it makes sense to define a renormalized Yukawa
coupling by $g_R^2 =Z_{\phi} Z_{\chi} g_0^2$, ($\alpha\equiv{{g_R^2}\over
{16\pi^2}}$). We also set $\lambda_0^{\prime} =\lambda_0 Z_{\phi}^2$ but,
unlike $\alpha$, do
not expect
$\lambda_0^{\prime}$ to stay finite when $\Lambda\rightarrow\infty$.
$Z_{\chi}$ and $Z_{\phi}$ are defined in terms of the bare couplings by
$f(\mu^2 )={{g(\mu^2 )}\over{\mu^2}}=1$.

After performing the angular part of the momentum integrals one gets
when the products in the saddle point equations for $x,y$ dependent quantities
are converted into convolutions in momentum space one finds:
\begin{eqnarray}
f(p^2 )=Z_{\chi}h_{\chi}(p^2 )+\sqrt{2}\alpha\bar\rho
\bigg [ {1\over 2}\int_{p^2}^{\infty}
{{dk^2 }\over{g(k^2 )}}\cr
+{1\over{p^2 }}\int_0^{p^2 }
{{k^2 dk^2 }\over{g(k^2 )}}-
{1\over{2p^4 }}\int_0^{p^2 }
{{k^4 dk^2 }\over{g(k^2 )}} \bigg ]\cr
g(p^2 )=Z_{\phi}p^2 h_{\phi}(p^2 )+{{\sqrt{2}\alpha}\over{\bar\rho}}
\bigg [ {{p^2}\over { 2}}\int_{p^2}^{\infty}
{{dk^2}\over{k^2 f(k^2 )}} \cr+
\int_0^{p^2} {{dk^2}\over{f(k^2 )}}-
{1\over{2p^2 }}\int_0^{p^2}{{k^2 dk^2}\over{f(k^2 )}} \bigg ] +m^2 \cr
m^2=Z_{\phi} m_0^2+{ {\lambda_0^{\prime}}\over{32\pi^2 }}
\int_0^{\infty} {{k^2 dk^2 }\over {g(k^2 ) }}\cr-
{{\sqrt{2}\alpha}\over{\bar\rho}}
\int_0^{\infty}{{dk^2 }\over {f(k^2 )}}
\label{eq:inteqs}
\end{eqnarray}
The massless case is obtained by adjusting $m_0^2$ so that $m^2 =0$ and then
it makes sense to define $h(p^2 )={{g(p^2 )}\over {p^2}}$ leading to
a set of equations of a more symmetric appearance
with an interesting bosonic--fermionic symmetry at $\bar\rho =1$.

Rescaling the functions and the momentum variable by
\begin{equation}
f(u\mu^2 )=\sqrt{2\alpha } \bar\phi( u)~~~~h(u\mu^2 )=
\sqrt{\alpha } \bar\psi( u)
\label{eq:resc}
\end{equation}
we obtain:
\begin{eqnarray}
\bar\psi (u) ={1\over{\sqrt{\alpha}}}+{{Z_{\phi}}\over{\sqrt{\alpha}}}
[h_{\phi} (u\mu^2 )-h_{\phi} (\mu^2 )]\cr+ {1\over{\bar\rho}}\int_u^1
{{dv}\over
{v}}
\int_0^1
{{(1-x)dx}\over{\bar\phi (xv)}}\cr
\bar\phi (u) ={1\over{2\sqrt{\alpha}}}+{{Z_{\chi}}\over{2\sqrt{\alpha}}}
[h_{\chi} (u\mu^2 )-h_{\chi} (\mu^2 )]\cr+ \bar\rho\int_u^1 {{dv}\over {v}}
\int_0^1
{{(1-x)dx}\over{\bar\psi (xv)}}
\label{eq:dblint}
\end{eqnarray}

It is clear that when $n$ is large enough $h$ and $f$ are dominated by
the $h_{\phi , \chi} $ functions in the ultraviolet and therefore the large
cutoff
behavior of the wave function renormalization constants is such
that in the infinite cutoff limit, at $u$ and $\mu^2$ fixed,
$Z_{\phi}$ and $Z_{\chi}$ simply disappear
from the above equation, leaving us with
\begin{eqnarray}
\bar\psi (u) ={1\over{\sqrt{\alpha}}}+
{1\over{\bar\rho}} \int_u^1 {{dv}\over {v}} \int_0^1
{{(1-x)dx}\over{\bar\phi (xv)}} \cr
\bar\phi (u) ={1\over{2\sqrt{\alpha}}}+
\bar\rho\int_u^1 {{dv}\over {v}} \int_0^1
{{(1-x)dx}\over{\bar\psi (xv)}}
\label{eq:inteqcont}
\end{eqnarray}

On physical grounds it is obvious that we wish that $\bar\psi$ and $\bar\phi$
be positive. The equations above show that if this is true in some interval
$(0,u_{*})$ then the functions will be monotonically decreasing there.
However, the equation
do not admit an asymptotic behavior as $u\rightarrow\infty$ with
both functions approaching non--negative limits so the positivity requirement
must get violated somewhere in the
ultraviolet.\footnote{A zero of $\bar\psi$ or $\bar\phi$ at a positive
$u$ corresponds to a pole in a two--point function
at an Euclidean momentum; the ``particle'' associated
with this pole would be tachyonic.} Starting at the scale where the
violation first occurs cutoff effects cannot be neglected any more. This
``unphysical'' scale is the usual Landau pole, this time appearing in a
nonperturbative approximation.

In the infrared there is no Landau pole problem and
cutoff effects are indeed negligible. To investigate the behavior there
we set $x=-\log (u)$ and $ \hat\psi (x) =\bar\psi (u),~
\hat\phi (x) =\bar\phi (u) $ and derive:
\begin{eqnarray}
{\cal D}\hat\psi ={1\over{\bar\rho\hat\phi}},~~~~
{\cal D}\hat\phi ={{\bar\rho}\over{\hat\psi}}\cr
{\cal D}\equiv{{d}\over{dx}}
\bigg ( {{d}\over{dx}} -1\bigg ) \bigg ({{d}\over{dx}} -2 \bigg )
\label{eq:diffeqs}
\end{eqnarray}

These equations allow us to extract the infrared behavior of the solution
of the integral equations. We ended up with only differential
equations (rather than integral)
reflecting that according to the Renormalization Group
one needs only very limited information at a given scale in order
to derive the behavior at a scale close by; therefore a differential
equation must show up eventually, its order less the number
of asymptotic conditions determining the
amount of information at a given scale that is needed to determine
the behavior at the next scale. Our choice of regularization was
made so that even at finite cutoff $\hat\phi$ and $\hat\psi$ obey
a set of purely differential equations of a structure similar
to~\ref{eq:diffeqs}.

Since the fixed point governing
the infrared behavior is the free field fixed point the form of the
solutions in the infrared simply embodies the two anomalous dimensions
associated with the $\chi$ and $\phi$ fields when
expanded in $\alpha$ around $\alpha =0$. These anomalous dimensions
as a function of the coupling also determine the $\beta$ function and
from the asymptotic series of $\hat\psi (x)$
and $\hat\phi (x)$ at $x\rightarrow\infty$ we can get the contributions
to the $\beta (\alpha )$ function ordered in the number of loops.

The differential equations lead to:
\begin{eqnarray}
\hat\psi\sim \psi_0 x^b [1+c{{\log x}\over x}+c^{\prime} {1\over x}+\cdots]\cr
\hat\phi\sim \phi_0 x^a [1+d{{\log x}\over x}+d^{\prime} {1\over x}+\cdots]
\label{eq:asympeq}
\end{eqnarray}
where
\begin{eqnarray}
a=b\bar\rho^2 ={{\bar\rho^2}\over{1+\bar\rho^2}},~
d=c\bar\rho^2 =-{{3\bar\rho^4}\over{(1+\bar\rho^2  )^3}}\cr
{{d^{\prime}}\over{\bar\rho^2 }}-c^{\prime} ={{3(1-\bar\rho^2 )}\over{
2(1+\bar\rho^2 )^2 }};~~~~~\psi_0\phi_0 ={{\bar\rho +\bar\rho^{-1}}\over 2}
\label{eq:expon}
\end{eqnarray}
Only the product $\psi_0\phi_0$ and a particular linear combination of
$d^{\prime}$ and $c^{\prime}$ get determined because
the equations are invariant
under a field rescaling $\hat\psi\rightarrow{A^{-1}}\hat\psi$,
$\hat\phi\rightarrow A \hat\phi$ reflecting a change in the finite parts of
the wave function renormalization constants and a
shift $x\rightarrow x+x_0$ reflecting a change in $\mu^2$. The
field--rescaling
invariance disappears in the product
$r(u)\equiv\sqrt{2} \bar\psi (u)\bar\phi (u)$
and this combination is indeed special because it satisfies a
Renormalization Group
equation in which only the $\beta$--function appears (the sum
of the two anomalous dimensions rather than each individual one):
\begin{equation}
\bigg [ -{{\partial}\over{\partial t}}+\beta (\alpha )
{{\partial}\over{\partial\alpha}} \bigg ] r(e^{2t} ,\alpha )=0
\label{eq:betadeff}
\end{equation}

\if \epsfpreprint Y
\begin{figure}[htb]
\epsfxsize=\columnwidth
\epsffile{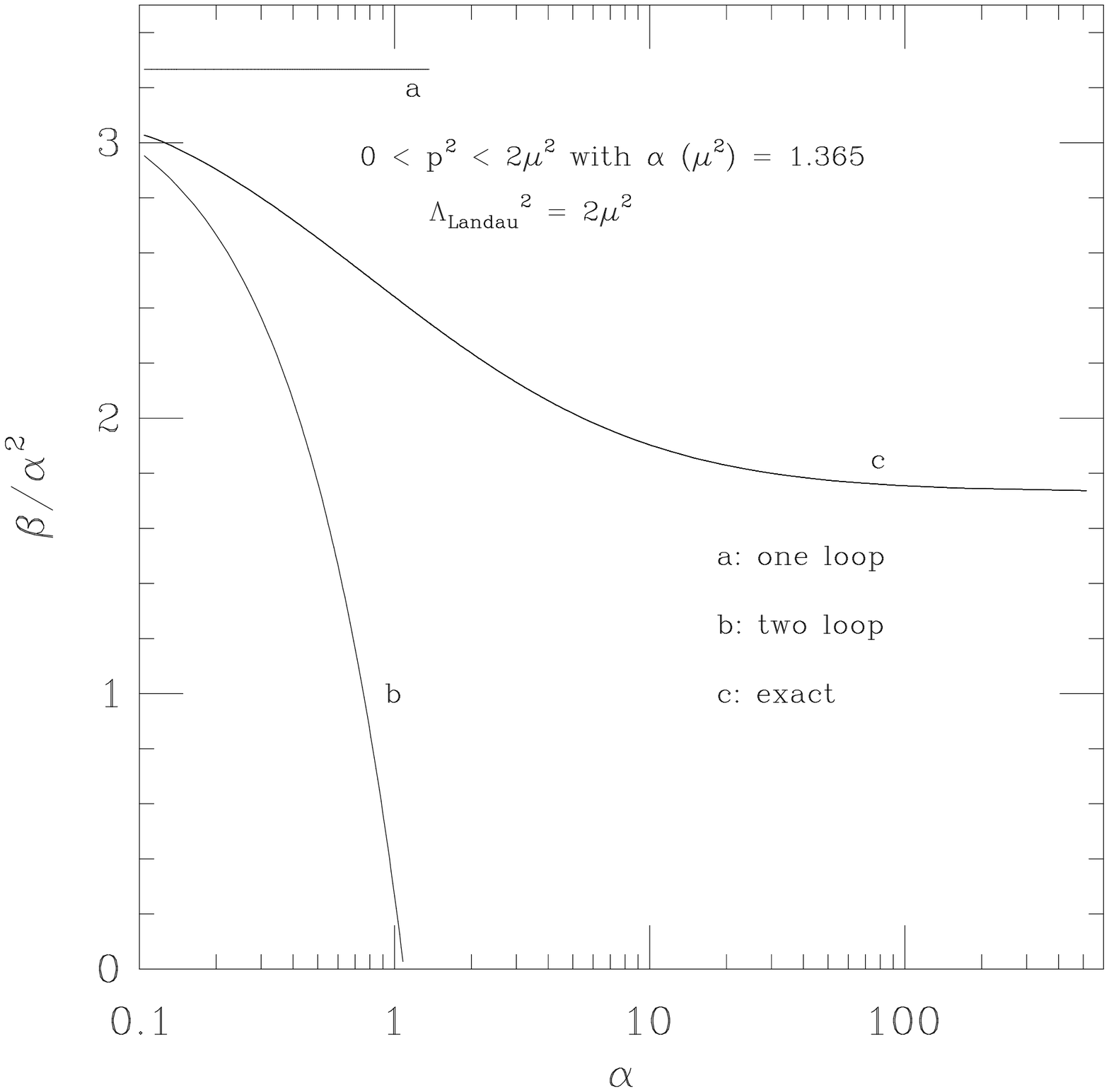}
\caption{\captionA}
\end{figure}
\fi

This equation is exact in our limit. From it we derive another exact relation
between $r$ and $\beta$:
\begin{equation}
\beta (\alpha (u\mu^2) )=
-2[\alpha (u\mu^2 )]^2
{{dr(u)}\over{d(\log u)}}{\bigg |}_{r(u)={1\over{\alpha (u\mu^2 )}}}
\label{eq:betaexact}
\end{equation}

Using now the asymptotic expansion of the exact
solution in the infrared we get:
\begin{equation}
\beta(\alpha )=\alpha^2 [\sqrt{2} (\bar\rho +\bar\rho^{-1})-3\alpha+O(\alpha^2
)]
\label{eq:betapert}
\end{equation}

The one loop result when expressed in a conventional variable, $y(t)$
(in the standard model the top mass at tree level is $m_t =y(\sim 0)~
246~[GeV]$),
is, with the true value $\bar\rho={1\over{\sqrt{3}}}$, ${{dy^2}\over
{dt}}={{y^4}\over{\pi^2}}$ which is close to the finite $N$ result,
${{dy^2}\over{dt}}={{9y^4}\over{8\pi^2}}$. The difference comes from the
wave function renormalization of the left handed $\psi$ field which is
suppressed at leading order in ${1\over N}$. For arbitrary $N_c$ and $N_w$
we would have obtained at one loop order
${{dy^2}\over{dt}}={{(2N_c +N_w +1) y^4}\over{8\pi^2}}$
and we see that the relative magnitude of the
$1/N$ correction is much smaller than in the scalar case\cite{dn}.

To get the full $\beta$--function at infinite $N$ we solve for $\hat\psi$
and $\hat\phi$ numerically by iterating an integral form of the
equations~\ref{eq:inteqs}
but with the normalization conditions at $\mu^2$
incorporated; this avoids the appearance of the
double integrals in~\ref{eq:dblint} and~\ref{eq:inteqcont} .

The result for $\beta$ is
shown in the figure and one again sees explicit evidence for the
nonperturbative existence of a Landau pole. We compare
the Landau pole energy obtained from the exact result to the one loop estimate.
For example, with the high value of $\alpha(\mu^2 )=1.365$ chosen in the figure
we obtain
$\Lambda_{\rm Landau}^2 =2\mu^2$ while at one loop
$\Lambda_{\rm Landau}^2 =1.565\mu^2$. A top quark corresponding
to such a strong coupling would have a mass of $m_t =4\pi\sqrt{{\alpha}
\over{12}}~ 246~[GeV]\approx 1043~GeV$. We see that ``triviality bounds''
on the top will likely come out close to the perturbative unitarity bounds of
Chanowitz et. al.\cite{chanfur} in contrast to
Einhorn and Goldberg\cite{einhgold} who obtained $5600 ~GeV$.

{}From our non--perturbative result
in the figure one sees that the one and two loop errors in $\beta$
are about comparable for $\alpha\approx 0.4$ (roughly
of order 20 percent) and that using the one loop results
all the way out to infinity is not very harmful, but the two loop
result would be completely misleading if extrapolated beyond its
region of validity, to $\alpha\approx 1$. A top mass of $200~GeV$ would
correspond to $\alpha\approx 0.05$, well within the perturbative domain.

For the future we plan the following steps:

$\bullet$ Compute the $\beta$--function associated with $\lambda$ at infinite
cutoff
in the massless case.

$\bullet$ Reintroduce the finite cutoff in the above computation to see the set
in and magnitude
of cutoff scaling violations.

$\bullet$  Redo both the infinite cutoff and the finite cutoff computation away
from the critical line,
i.e. when thresholds are present. This has to be done in both the broken and
the symmetric phases.

$\bullet$  Compute the effective potential at infinite cutoff limit first and
then at finite cutoff and analyze
the issue of  ``vacuum instability''  in this non--perturbative setting.

$\bullet$  Study bound state problems.

\if \epsfpreprint N
\section{Figure captions}

\noindent{\bf Figure~$1$:~}\captionA

\fi

\vfill

\end{document}